\documentclass[epj]{svjour}

\usepackage{amsmath,amssymb}
\usepackage{graphics}
\usepackage{upgreek}

\newcommand{\Cc}{C^{({\rm c})}}
\newcommand{\Cp}{C^{(p)}}
\newcommand{\Cr}{C^{({\rm r})}}
\newcommand{\eg}{{{\it e.g.}, }}
\newcommand{\etab}{\eta_{\rm b}}
\newcommand{\etal}{\eta_l}
\newcommand{\fc}{f^{({\rm c})}}
\newcommand{\fr}{f^{({\rm r})}}
\newcommand{\fp}{f^{(p)}}
\newcommand{\Gb}{G_{\rm b}}
\newcommand{\half}{\frac{1}{2}}
\newcommand{\ie}{{{\it i.e.}, }}

\newcommand{\lc}{\ell_{\rm c}}
\newcommand{\pd}{\partial}

\newcommand{\rc}{r_{\rm c}}
\newcommand{\rhat}{{\bf\hat{r}}}
\newcommand{\rmd}{{\rm d}}
\newcommand{\thetahat}{\bf\hat{\boldsymbol{\uptheta}}}
\newcommand{\vecf}{{\bf f}}
\newcommand{\vecF}{{\bf F}}
\newcommand{\vecr}{{\bf r}}
\newcommand{\vecu}{{\bf u}}
\newcommand{\vecU}{{\bf U}}
\newcommand{\vecv}{{\bf v}}
\newcommand{\vecV}{{\bf V}}
\newcommand{\vecVc}{{\bf V}^{\rm (c)}}
\newcommand{\vecVr}{{\bf V}^{\rm (r)}}

\newcommand{\zhat}{\hat{\bf z}}

\begin{document}

\title{Response of a polymer network to the motion of a rigid sphere}

\author{Haim Diamant
}


\institute{Raymond \& Beverly Sackler School of
    Chemistry, Tel Aviv University, Tel Aviv 6997801, Israel 
}

\date{Received: date / Revised version: date}

\abstract{In view of recent microrheology experiments we re-examine
  the problem of a rigid sphere oscillating inside a dilute polymer
  network. The network and its solvent are treated using the two-fluid
  model. We show that the dynamics of the medium can be decomposed
  into two independent incompressible flows. The first, dominant at
  large distances and obeying the Stokes equation, corresponds to the
  collective flow of the two components as a whole. The other,
  governing the dynamics over an intermediate range of distances and
  following the Brinkman equation, describes the flow of the network
  and solvent relative to one another. The crossover between these two
  regions occurs at a dynamic length scale which is much larger than
  the network's mesh size. The analysis focuses on the spatial
  structure of the medium's response and the role played by the
  dynamic crossover length. We examine different boundary conditions
  at the sphere surface. The large-distance collective flow is shown
  to be independent of boundary conditions and network
  compressibility, establishing the robustness of two-point
  microrheology at large separations. The boundary conditions that fit
  the experimental results for inert spheres in entangled F-actin
  networks are those of a free network, which does not interact
  directly with the sphere. Closed-form expressions and scaling
  relations are derived, allowing for the extraction of material
  parameters from a combination of one- and two-point
  microrheology. We discuss a basic deficiency of the two-fluid model
  and a way to bypass it when analyzing microrheological data.
}

\maketitle

\section{Introduction}
\label{sec_intro}


In the past two decades the technique of microrheology has been used
to characterize the dynamic response of soft and biological matter
\cite{Squires2010}. In one-point microrheology \cite{Mason1995} the
viscoelastic moduli of the material are inferred from the
displacements of a tracer particle in response to an external force
(active microrheology), or its displacement autocorrelations under
thermal fluctuations (passive microrheology). Such measurements rely
on a generalized Stokes relation (GSR, active) or generalized
Stokes-Einstein relation (GSER, passive), asserting that the
particle's response has the same form as in a viscous fluid, with the
fluid's shear viscosity $\eta$ replaced by
$\etab(\omega)=\Gb(\omega)/(i\omega)$. Here $\Gb(\omega)$ is the
frequency-dependent complex shear modulus of the bulk
material. Subsequently, two-point microrheology was introduced as well
\cite{Crocker2000}. In this technique the moduli are deduced from the
displacements of one particle in response to a force exerted on
another (active), or the displacement cross-correlations of
fluctuating particle pairs (passive), as a function of their mutual
distance. The two-point measurements are based on a ``generalized
Oseen tensor'', assuming that the spatial response at large distances
is the same as in a viscous fluid, with the aforementioned
replacement. Although it is considered more reliable than the
one-point technique, two-point microrheology has not been used as
widely, mainly because of the difficulty to accumulate enough
statistics for particle pairs at each given separation.

When results from one-point microrheology are compared with
macrorheology and two-point microrheology, discrepancies are found for
various materials
\cite{Crocker2000,McGrath2000,Chen2003,Gardel2003,Starrs2003,Valentine2004}.
One-point measurements commonly yield much smaller moduli than the
other two techniques. The disagreement has been attributed to
differences between the particle's immediate environment and the bulk
material, leading to deviations from the GSR/GSER. Indeed, one-point
measurements were found to be sensitive to the surface chemistry of
the tracer particle \cite{McGrath2000,Valentine2004}, whereas
two-point measurements were not \cite{Valentine2004}. Theoretical
attempts to account for such local effects have included modifying the
boundary conditions at the particle surface between no slip, partial
slip, and full slip \cite{Starrs2003,Fu2008}, and the introduction of
a shell of different viscoelastic properties surrounding the particle
\cite{Levine2000,Levine2001b}.

A recent study, applying two-point microrheology to entangled F-actin
networks \cite{prl14,SonnSegev14}, revealed a wide range of distances,
intermediate between the microscopic scale (the network's mesh size
$\xi$) and the macroscopic (asymptotically large) one, in which the
dynamic pair correlations were qualitatively different from those at
larger distances. This intermediate behavior ended at a distinct
dynamic length, $\lc$, much larger than $\xi$, which marked the
crossover to the macroscopic response. Theoretical arguments given in
ref.\ \cite{prl14} showed that the intermediate response was
fundamentally different from the macroscopic one, in that it was
related to mass transport, rather than momentum transport, of the
fluid. As we shall see below, the intermediate region corresponds to
the relative motion of the material's two components (network and
solvent), whereas in the asymptotic region the material moves
collectively, as a whole. The application of these ideas to a
combination of one- and two-point measurements gives rise to an
extension of microrheology, allowing for better characterization of
complex fluids (\eg extracting their correlation length)
\cite{prl14,SonnSegev14}.

In the present article we elaborate on, and extend, the theoretical
results briefly presented in ref.\ \cite{prl14}. We begin in
sect.\ \ref{sec_problem} by defining the problem, which is the
analogue of Stokes' problem for the motion of a rigid sphere, with the
viscous fluid replaced by a two-fluid medium
\cite{deGennes1976a,deGennes1976b,Doi1992,Milner1993,Levine2001a,Bruinsma2014}.
An approximate treatment of the dynamics of a bead embedded in such a
medium was presented by Levine and Lubensky \cite{Levine2001a}. The
exact solution of the Stokes-analogous problem has already been
derived by Fu {\it et al.}  \cite{Fu2008}. In sect.\ \ref{sec_general}
we solve it again from a slightly different perspective, providing
additional physical insight. We derive the general solution and
extract from it several general properties, which are independent of
boundary conditions. In sect.\ \ref{sec_particular} we present the
particular solutions for three different sets of boundary
conditions. The first two, corresponding to no slip and full slip of
the network over the sphere surface, were treated in
ref.\ \cite{Fu2008} as limiting cases of a general slip condition. For
these cases the present analysis offers expressions of different
experimental utility. The third set of boundary conditions, which
describes a network having no direct interaction with the sphere, has
not been treated before. The properties of this particular solution
are substantially different from the other two. We show that they match
the experimental observations for entangled F-actin networks. In
sect.\ \ref{sec_particular} we also derive scaling relations, which
can facilitate the analysis of experimental data and the extraction of
material parameters from them. Section~\ref{sec_discuss} discusses in
detail the results and their implications.

The analyses in
refs.\ \cite{Fu2008,Levine2000,Levine2001b,Levine2001a} focused on the
response of the sphere as a function of frequency. We emphasize the
spatial response of the medium, its relation to basic conservation
laws, and its dependencies on distance, sphere radius, and the
material's characteristic lengths. We highlight, in particular, the
key role played by the dynamic crossover length $\lc$.

\section{The problem}
\label{sec_problem}

\begin{figure}[tbh]
\centerline{\resizebox{0.48\textwidth}{!}
{\includegraphics{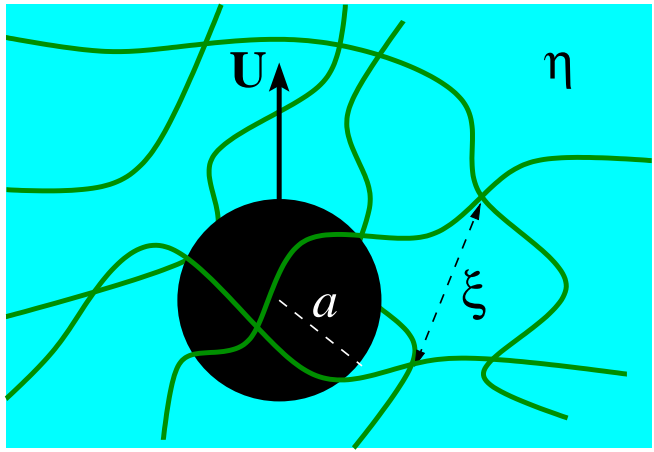}}}
\vspace{1cm}
\caption[]{Schematic view of the system.}
\label{fig_scheme}
\end{figure}

Figure \ref{fig_scheme} shows a schematic view of the system. A rigid
sphere of radius $a$ moves with velocity $U$ inside a polymer network.
As done in earlier studies, we use the two-fluid model
\cite{deGennes1976a,deGennes1976b,Doi1992,Milner1993,Levine2001a,Bruinsma2014}
to describe the medium. The main advantage of this model is that it is
sufficiently simple to be treated analytically while delivering the
key features of a complex fluid\,---\,emergent correlation length
$\xi$ and bulk viscoelastic modulus $\Gb(\omega)$. Its main
disadvantage is that it is a continuous, linear, hydrodynamic model,
neglecting effects of thermal fluctuations, small-scale
heterogeneities, and nonlinear advection. An additional fundamental
shortcoming of the model will be discussed in
sect.\ \ref{sec_discuss}.

In the presentation below all fields are position- and time-dependent,
and we Fourier-transform them from the time to the frequency domain,\\
$f(\vecr,t)\rightarrow
f(\vecr,\omega)=\int_{-\infty}^{\infty}\rmd t e^{-i\omega t}f(\vecr,t)$.

The model has two components. The first is a (visco)elastic network,
having a volume fraction field $\phi_u(\vecr,\omega)$, displacement
field $\vecu(\vecr,\omega)$, and the stress tensor
\begin{equation}
  \sigma^{(u)}_{ij} = 2G [u_{ij} - (u_{kk}/3)\delta_{ij}] + K u_{kk}\delta_{ij},
\label{sigmaelastic}
\end{equation}
where $u_{ij}\equiv\half(\pd_iu_j+\pd_ju_i)$ is the network's strain
tensor, and $G$ and $K$ its shear and compression moduli, which may be
frequency-dependent. The second component is a viscous fluid, having a
volume fraction field $\phi_v(\vecr,\omega)$, flow velocity field
$\vecv(\vecr,\omega)$, pressure field $p(\vecr,\omega)$, and the
stress tensor
\begin{equation}
  \sigma^{(v)}_{ij} = -p\delta_{ij} + 2\eta [v_{ij} -
    (v_{kk}/3)\delta_{ij}] + \zeta v_{kk}\delta_{ij},
\label{sigmaviscous}
\end{equation}
where $v_{ij}\equiv\half(\pd_iv_j+\pd_jv_i)$ is the fluid's
strain-rate tensor, and $\eta$ and $\zeta$ are its shear and
compression viscosities. The two components are coupled via mutual
friction characterized by a coefficient $\Gamma$. The five governing
equations for the five fields ($\vecu,\vecv,p,\phi_u,\phi_v$) are as
follows:
\begin{eqnarray}
  -\omega^2\rho_u\vecu &=& \nabla\cdot\boldsymbol\sigma^{(u)}
  - \Gamma(i\omega\vecu-\vecv) + \vecf_u \nonumber\\
  &=& G\nabla^2\vecu + (K+G/3)\nabla(\nabla\cdot\vecu) \nonumber\\
  && - \Gamma(i\omega\vecu-\vecv) + \vecf_u \\
  i\omega\rho_v\vecv &=& \nabla\cdot\boldsymbol\sigma^{(v)} 
  - \Gamma(\vecv-i\omega\vecu) + \vecf_v \nonumber\\
  &=& -\nabla p + \eta\nabla^2\vecv +
  (\zeta+\eta/3)\nabla(\nabla\cdot\vecv) \nonumber\\
  && - \Gamma(\vecv-i\omega\vecu) + \vecf_v\\ 
  0 &=& \phi_u + \nabla\cdot(\phi_u\vecu) \\
  0 &=& i\omega\phi_v + \nabla\cdot(\phi_v\vecv) \\
  1 &=& \phi_u + \phi_v,
\end{eqnarray}
where $\rho_u,\rho_v$ are the mass densities of the two components,
and $\vecf_u(\vecr,\omega),\vecf_v(\vecr,\omega)$ are external force
densities exerted on them. The first two equations, together, reflect
the conservation of momentum in the composite material; the third and
fourth\,---\,the conservation of mass of each component separately;
and the last\,---\,the assumption of incompressibility for the
composite.

We employ the following simplifications. (i) Inertial effects are
omitted; they are negligible at sufficiently low frequencies and can
easily be included if needed \cite{Levine2001a}. (ii) The network is
taken as semidilute, $\phi_u\ll\phi_v\simeq 1$. (iii) We specialize to
the case of no external forces, $\vecf_u=\vecf_v=0$. Under these
assumptions the governing equations attain the much simpler form,
\begin{eqnarray}
\label{vecu}
  0 &=& G\nabla^2\vecu + (K+G/3)\nabla(\nabla\cdot\vecu)
  - \Gamma(i\omega\vecu-\vecv) \\
\label{vecv}
  0 &=& -\nabla p + \eta\nabla^2\vecv
  - \Gamma(\vecv-i\omega\vecu) \\ 
\label{incompress}
  0 &=& \nabla\cdot\vecv.
\end{eqnarray}

The perturbation that we introduce is in the form of a rigid sphere,
centered at the origin and moving in the $z$ direction with velocity
$\vecU=U(\omega)\zhat$. Under these conditions the problem has
azimuthal symmetry. Hence, using spherical coordinates
$(r,\theta,\phi)$, all the fields depend on $r$ and $\theta$ alone,
and the vector fields have only $r$ and $\theta$
components. Consequently, the stresses of eqs.\ (\ref{sigmaelastic})
and (\ref{sigmaviscous}) take the form,
\begin{eqnarray}
  \sigma^{(u)}_{rr} &=& K \left[ \frac{1}{r^2} \pd_r(r^2u_r) + 
  \frac{1}{r\sin\theta} \pd_\theta(u_\theta \sin\theta) \right]
 \nonumber\\
  &&+ \frac{2G}{3} \left[ 2r\pd_r(u_r/r) - 
  \frac{1}{r\sin\theta} \pd_\theta(u_\theta \sin\theta) \right]
 \nonumber\\
  \sigma^{(u)}_{r\theta} &=& G\left[ r\pd_r(u_\theta/r) + \frac{1}{r}
  \pd_\theta u_r \right]
 \nonumber\\
  \sigma^{(v)}_{rr} &=& -p + 2\eta \pd_rv_r
 \nonumber\\
  \sigma^{(v)}_{r\theta} &=& \eta\left[ r\pd_r(v_\theta/r) + \frac{1}{r}
  \pd_\theta v_r \right].
\label{sigmapolar}
\end{eqnarray}

We consider an unbounded medium, which is unperturbed and stationary
far away from the sphere. The boundary conditions at infinity are,
therefore,
\begin{equation}
  \vecu(r\rightarrow\infty,\theta) = 
  \vecv(r\rightarrow\infty,\theta) = 
  \nabla p(r\rightarrow\infty,\theta) = 0.
\label{bcinfinity}
\end{equation}
As to the boundary conditions at the surface of the sphere, we will
examine various possibilities in sect.\ \ref{sec_particular}.

\section{General solution}
\label{sec_general}

\subsection{Decoupled flows}

First, we take the divergence of eq.\ (\ref{vecv}) and use
eq.\ (\ref{incompress}) to obtain a relation between the compressive
stresses of the two components,
\begin{equation}
  \nabla\cdot\vecu = \frac{1}{i\omega\Gamma} \nabla^2p.
\end{equation}

The next step is to decouple the equations for the two vector fields,
$\vecu$ and $\vecv$. We apply the following linear transformation:
\begin{eqnarray}
 \label{Vcdef}
  \vecVc &=& \left(1 - \frac{\eta}{\etab}\right)
   \left( i\omega\vecu + \frac{\eta}{\etab-\eta}\vecv 
   - \frac{1}{\Gamma} \nabla p \right)\\
 \label{Vrdef}
  \vecVr &=& \left(\frac{\etab}{\eta} - 1\right)
    \left( i\omega\vecu - \vecv - \frac{1}{\Gamma}\nabla p \right)\\
 \label{Pdef}
  P &=& p - \lambda^2\nabla^2 p,
\end{eqnarray}
whereupon eqs.\ (\ref{vecu})--(\ref{incompress}) turn into
\begin{eqnarray}
 \label{vecVc}
  0 &=& -\nabla P + \etab \nabla^2\vecVc \\
 \label{vecVr}
  0 &=& -\nabla P + \eta \left( \nabla^2\vecVr - \xi^{-2}\vecVr \right)\\
 \label{incompress2}
  0 &=& \nabla\cdot\vecVc = \nabla\cdot\vecVr.
\end{eqnarray}
In these equations we have introduced the following bulk viscosity and
characteristic lengths:
\begin{eqnarray}
 \label{etabdef}
  \etab &\equiv& G/(i\omega) + \eta\\
 \label{xidef}
  \xi &\equiv&  \left( \frac{G\eta}{i\omega\Gamma\etab} \right)^{1/2} \\
 \label{lambdadef}
  \lambda &\equiv& \left( \frac{K+4G/3}{i\omega\Gamma} \right)^{1/2} = 
  \left[ \frac{2(1-\nu)}{1-2\nu} \frac{\etab}{\eta} \right]^{1/2} \xi,
\end{eqnarray}
where $\nu$ is the network's Poisson ratio.

Thus, the dynamics of the two-fluid model
[eqs.\ (\ref{vecu})--(\ref{incompress})] have been decomposed into two
independent incompressible flows, $\vecVc$, and $\vecVr$. The former,
in which $\vecv$ and $i\omega\vecu$ appear with the same sign,
describes the collective flow of the network--fluid composite. This
flow satisfies the scale-free Stokes equation (\ref{vecVc}) with
viscosity given by the bulk viscosity $\etab(\omega)$. The latter,
containing $\vecv$ and $i\omega\vecu$ with opposite signs,
corresponds to the relative flow of the two components. It obeys the
Brinkman equation for a fluid embedded in a porous medium
\cite{Brinkman}, eq.\ (\ref{vecVr}), depending on the fluid viscosity
$\eta$ and characteristic pore size $\xi$. Hence, $\xi$ is identified
with the network's mesh size \cite{Levine2000}, up to a
proportionality factor close to unity \cite{prl14,SonnSegev14}. It
also characterizes the spatial decay of transverse (shear) stresses
due to the friction between the two components and, therefore,
decreases with increasing $\Gamma$. The second length, $\lambda$,
characterizes the decay of longitudinal (compressive) stresses and
diverges in the limit of an incompressible network
($K\rightarrow\infty$ or $\nu=1/2$).  In addition, we define another
dynamic length,
\begin{equation}
  \lc \equiv \left(\frac{\etab}{\eta}\right)^{1/2} \xi.
\label{lcdef}
\end{equation}
As we shall see, $\lc$ plays a crucial role in the dynamics of the
medium. It is proportional to $\lambda$ but remains finite in the
limit of an incompressible network. We note the hierarchy
$\xi<\lc<\lambda$. Usually (\eg for sufficiently low frequencies),
$\etab\simeq G/(i\omega)\gg\eta$, and then
$\xi\ll\lc<\lambda$. Additionally, for a network close to the
incompressible limit, we have the complete separation of scales
$\xi\ll\lc\ll\lambda$.

\subsection{Stream functions and resulting solution}

Following Stokes' original scheme, we express the vector flows in
terms of scalar stream functions,
\begin{equation}
  \vecV^{({\rm c,r})} = -\frac{1}{r^2\sin\theta} \pd_\theta
  \psi^{({\rm c,r})} \rhat + 
  \frac{1}{r\sin\theta} \pd_r\psi^{({\rm c,r})} \thetahat,
\label{psicr}
\end{equation}
and assume the following separation of variables:
\begin{eqnarray}
  \psi^{({\rm c,r})}(r,\theta) &=& f^{({\rm c,r})}(r) \sin^2\theta 
\label{fcr}\\
  p(r,\theta) &=& \fp(r) \cos\theta.
\label{fp}
\end{eqnarray}

Substituting eqs.\ (\ref{psicr}) and (\ref{fcr}) in eq.\ (\ref{vecVc})
(after taking the curl of the latter to eliminate $\nabla P$), we get
the equation for $f^{({\rm c})}$,
\begin{equation}
  0 = r^4(\fc)'''' - 4r^2(\fc)'' + 8r(\fc)' - 8\fc.
\label{fceq}
\end{equation}
Its solution is
\begin{equation}
  \fc(r) = \Cc_1 r + \frac{\Cc_2}{r},
\label{fcsol}
\end{equation}
where we have omitted terms which do not decay to zero after division
by $r^2$, to ensure a vanishing $\vecVc$ at infinity. The resulting
collective flow is
\begin{equation}
  \vecVc = -2 \left( \frac{\Cc_1}{r} + \frac{\Cc_2}{r^3} \right)
  \cos\theta\rhat
  + \left( \frac{\Cc_1}{r} - \frac{\Cc_2}{r^3} \right)
  \sin\theta\thetahat.
\label{vecVcsol}
\end{equation}
The collective flow contains the $1/r$ and $1/r^3$ terms known from
azimuthally symmetric solutions of the Stokes equation, the first
arising from the transverse component of a momentum monopole, and the
second from a combination of a momentum quadrupole and a mass dipole
\cite{prl14}.

Repeating the same procedure for the relative flow, we get,
instead of eqs.\ (\ref{fceq})--(\ref{vecVcsol}),
\begin{eqnarray}
  0 &=& r^4\xi^2(\fr)'''' - r^2(r^2+4\xi^2)(\fr)''  
  \nonumber\\ &&+\; 8r\xi^2(\fr)' + 2(r^2-4\xi^2)\fr\\
  \fr(r) &=& \frac{\Cr_1}{r} + \Cr_2(1+\xi/r)e^{-r/\xi}\\
  \vecVr &=& -2 \left( \frac{\Cr_1}{r^3} + \frac{\Cr_2(r+\xi)}{r^3}
  e^{-r/\xi} \right) \cos\theta \rhat \nonumber\\
  &&-\left( \frac{\Cr_1}{r^3} + \frac{\Cr_2(r^2+\xi r+\xi^2)}{\xi r^3} 
  e^{-r/\xi}\right) \sin\theta\thetahat.
\label{vecVrsol} \nonumber\\
  && 
\end{eqnarray}
The relative flow contains the $1/r^3$ and exponentially small terms
known from solutions of the Brinkman equation, the former arising from
a mass dipole, and the latter from the spatially decaying transverse
momentum, whose decay length is $\xi$ \cite{LongAjdari2001,ijc07,jpsj09}.

We substitute eqs.\ (\ref{Pdef}), (\ref{fp}), and (\ref{vecVcsol}) in
eq.\ (\ref{vecVc}) to find the equation for $\fp$,
\begin{equation}
  0 = \lambda^2r^2(\fp)'' + 2\lambda^2r(\fp)' - 
  (r^2+2\lambda^2)\fp - 2\etab\Cc_1,
\end{equation}
whose solution is
\begin{equation}
  \fp(r) = -\frac{2\etab\Cc_1}{r^2} + 
  \frac{\Cp(r+\lambda)}{r^2} e^{-r/\lambda},
\end{equation}
where we have omitted a term that diverges at infinity. The resulting
pressure is 
\begin{equation}
  p = \left( -\frac{2\etab\Cc_1}{r^2} + 
  \frac{\Cp(r+\lambda)}{r^2} e^{-r/\lambda} \right) \cos\theta.
\label{psol}
\end{equation}
It contains a $1/r^2$ term, arising from the longitudinal component of
a momentum monopole, and exponentially small terms, corresponding to
spatially decaying longitudinal stresses whose decay length is
$\lambda$. The transformed pressure $P$ [eq.\ (\ref{Pdef})] contains
only the $1/r^2$ term.

We can use eq.\ (\ref{vecVr}) for the relative flow, rather than
eq.\ (\ref{vecVc}) for the collective one, and obtain another
expression for $\fp$. Equating the two expressions gives a relation
between $\Cr_1$ and $\Cc_1$,
\begin{equation}
  \Cr_1 = (2\etab\xi^2/\eta) \Cc_1 = 2\lc^2 \Cc_1.
\label{Cr1Cc1}
\end{equation}
This is the first explicit appearance of the dynamic length $\lc$,
relating the leading terms in the collective and relative flows.

The general solution that is regular at infinity contains four
integration constants $(\Cc_1,\Cc_2,\Cr_2,\Cp)$ to be determined in
sect.\ \ref{sec_particular} from boundary conditions. However, we
first present several quantities of interest in terms of the
undetermined constants. We use them to derive properties stemming from
the fundamental response of the medium\,---\,in fact, any isotropic
viscoelastic medium\,---\, independent of specific boundary
conditions.

\subsection{Properties of the general solution}

The general solution for $\vecV^{({\rm c,r})}$, eqs.\ (\ref{vecVcsol})
and (\ref{vecVrsol}), can be substituted back in eqs.\ (\ref{Vcdef})
and (\ref{Vrdef}) to obtain the original fields, $\vecu(r,\theta)$ and
$\vecv(r,\theta)$. The resulting expressions can be found in the
Supplementary Material \cite{supplementary}.

The first quantity of interest is the total force $\vecF$ exerted on
the sphere. We substitute the general solution for $\vecu$ and $\vecv$
in the stress tensors, eq.\ (\ref{sigmapolar}), and integrate the
stresses over the surface of the sphere to obtain the force. The
separate network and fluid contributions to the force are found in
ref.\  \cite{supplementary}. Once they are added together,
most terms vanish, leaving
\begin{equation}
  \vecF = -8\pi\etab\Cc_1 \zhat.
\label{Ftot}
\end{equation}
Recall that $\Cc_1$ is associated with the long-range $1/r$ term in
the collective flow, eq.\ (\ref{vecVcsol}), and the corresponding
long-range $1/r^2$ term in the pressure, eq.\ (\ref{psol}). The simple
form of eq.\ (\ref{Ftot}) stems from momentum conservation\,---\,at
large distances, only the $1/r$ collective flow is at play, and it
corresponds to the flow due to a momentum monopole $\vecF$, with
effective viscosity $\etab$.

Let us examine the large-distance behavior of the flows in more
detail. The asymptotic term in eq. (\ref{vecVcsol}), as well as those
extracted from the expressions for $\vecu$ and $\vecv$ are
\begin{equation}
  \vecV^{\rm (c,0)} = \vecv^{(0)} = i\omega\vecu^{(0)} = 
  -\frac{\Cc_1}{r} \left(
  2\cos\theta\rhat - \sin\theta\thetahat \right),
\label{asymp}
\end{equation}
whereas $\vecV^{\rm (r,0)}=0$. Thus, far away from the perturbation
the two components flow together.  Combining eqs.\ (\ref{Ftot}) and
(\ref{asymp}), we find that, in terms of the force $\vecF$, the
large-distance flow has a universal form independent of boundary
conditions, mesh size, and compressibility,
\begin{equation}
  \vecV^{\rm (c,0)} = \vecv^{(0)} = i\omega\vecu^{(0)} =
  \frac{1}{8\pi\etab r} \left( 2 F_r\rhat + F_{\theta} \thetahat
  \right).
\label{Oseen}
\end{equation}
It represents a generalized Oseen tensor, with $\etab$ replacing
$\eta$, which is the basis of the original two-point microrheology
\cite{Crocker2000}. Its universality stems, once again, from the
conservation at large distance of transverse momentum emanating from
the source $\vecF$. The Oseen tensor describes a longitudinal flow
response (relation between the $r$-components of the force and
velocity) and a transverse response (relation between the
$\theta$-components), which are both positive. That is, if a second
particle is placed along or perpendicular to the direction of the
sphere's motion, in both cases it will be advected in the same
direction. The general result expressed in eq.\ (\ref{Oseen})
disagrees with earlier theoretical ones concerning the far-field flow
\cite{Fu2008,Levine2000,Levine2001b}. In
refs. \cite{Levine2000,Levine2001b} the far field was found to depend
on network compressibility. (Indeed, it was suggested as a means to
extract the compressibility from two-point correlations.) In
ref.\ \cite{Fu2008} it was shown to depend on network compressibility,
as well as the specific choice of boundary conditions, thus
undermining the universality of asymptotic two-point
microrheology. The apparent contradictions between these results and
ours are discussed, and resolved, in sect.\ \ref{sec_discuss}.

We are interested also in the sub-asymptotic $1/r^3$ terms,
\begin{eqnarray}
  \vecV^{\rm (c,1)} &=& -\frac{\Cc_2}{r^3} \left(
  2\cos\theta\rhat + \sin\theta\thetahat \right) \nonumber\\
  \vecV^{\rm (r,1)} &=& -\frac{2\lc^2\Cc_1}{r^3} \left(
  2\cos\theta\rhat + \sin\theta\thetahat \right),
\label{subasympV}
\end{eqnarray}
where we have used eq.\ (\ref{Cr1Cc1}) to replace $\Cr_1$ with
$\Cc_1$. The sub-asymptotic terms describe also relative motion of the
two components. The terms in the collective and relative flows depend
on different coefficients. In the case of the relative flow, we can
use eq.\ (\ref{Ftot}) to substitute $\Cc_1$ in eq.\ (\ref{subasympV})
and obtain a universal form in terms of $\vecF$,
\begin{equation}
  \vecV^{\rm (r,1)} = \frac{\lc^2}{4\pi\etab r^3} \left( 2 F_r\rhat -
  F_{\theta} \thetahat \right).
\label{subasympVr}
\end{equation}
This expression describes the flow created by a mass dipole equal to
$-(\lc^2/\etab)\vecF = -(\xi^2/\eta)\vecF$. It is the large-distance
flow caused by a point force in a Brinkman fluid
\cite{LongAjdari2001,ijc07,jpsj09}. One of its footprints is its sign,
implying a positive effect on the longitudinal response and a negative
effect on the transverse one.  Comparing eqs.\ (\ref{Oseen}) and
(\ref{subasympVr}), we identify $\sqrt{2}\,\lc$ as the distance at
which the leading terms in the collective and relative flows become
equal.

The original flow fields, which are more accessible experimentally,
contain combinations of sub-asymptotic terms from both the collective
and relative flows,
\begin{eqnarray}
\label{subasymp}
  \vecv^{(1)} &=& -\frac{\Cc_2-2\xi^2\Cc_1}{r^3}
  \left(2\cos\theta\rhat + \sin\theta\thetahat\right) \\
  i\omega\vecu^{(1)} &=& -\frac{\Cc_2-2(\lc^2+\xi^2)\Cc_1}{r^3}
  \left(2\cos\theta\rhat + \sin\theta\thetahat\right),
 \nonumber
\end{eqnarray}
such that $\vecv^{(1)}-i\omega\vecu^{(1)} = \vecV^{\rm (r,1)}$, the
universal expression of eq.\ (\ref{subasympVr}).  Because of these
combinations, the observed distances of crossover between the
sub-asymptotic and asymptotic behaviors do not necessarily coincide
with $\lc$. Hence, we define two such distances, $\rc^{(v)}$ and
$\rc^{(u)}$, as those at which $|\vecv^{(0)}|=|\vecv^{(1)}|$ and
$|\vecu^{(0)}|=|\vecu^{(1)}|$, respectively. According to
eqs. (\ref{asymp}) and (\ref{subasymp}) this occurs for
\begin{equation}
  \rc^{(v)} = [(\rc^{(u)})^2+2\lc^2]^{1/2} = | \Cc_2/\Cc_1 - 2\xi^2 |^{1/2}.
\label{rc}
\end{equation}

\section{Particular solutions for different boundary conditions}
\label{sec_particular}

The general solution obtained in the preceding section, after imposing
the boundary conditions at infinity, contains four integration
constants, $(\Cc_1,\Cc_2,\Cr_2,\Cp)$. Four boundary conditions are
required, therefore, at the sphere surface. In the following we
consider three cases for these boundary conditions. They correspond to
different physical conditions for the interaction between the sphere
and the network, as will be discussed in
sect.\ \ref{sec_discuss}. Other cases can be readily studied using the
file provided as Supplementary Material \cite{supplementary}.

\subsection{Sticking fluid and network}
\label{sec_bc1}

The first case that we study is stick boundary conditions for both
network and fluid, \ie the velocities of the two components at the
sphere surface are equal to the velocity of the sphere,
\begin{equation}
  \vecv(a,\theta) = i\omega\vecu(a,\theta) = U\zhat.
\label{bc1}
\end{equation}
We substitute these four conditions in the expressions for $\vecu$ and
$\vecv$, and solve for the four constants \cite{supplementary}.

Using eq.\ (\ref{Ftot}), we find for the total friction coefficient, 
$\gamma\equiv F/U$,
\begin{equation}
  \frac{\gamma}{6\pi\etab a} = 
  1 - \frac{\lc^2-\xi^2} {a^2+2\lambda^2+2\lambda a+\xi a+\lc^2}.
\label{gamma1}
\end{equation}
Recalling the hierarchy $\xi<\lc<\lambda$, we find that deviations
from the GSR can only be negative, and we also identify two relevant
limits. For a very large sphere,
\begin{equation}
  a\gg\lc:\ \ 
  \gamma \simeq 6\pi\etab a, 
\label{gamma1largea}
\end{equation}
$\gamma$ obeys, as expected, the generalized Stokes relation (GSR)
with viscosity $\etab(\omega)$. The condition for
eq.\ (\ref{gamma1largea}) to hold, however, is harder to fulfill than
what is usually assumed, since $\lc$ may be much larger than the mesh
size $\xi$. For smaller spheres,
\begin{equation}
  a\ll\lc:\ \ \frac{\gamma}{6\pi\etab a} \simeq
  1 - \frac{\lc^2}{2\lambda^2+\lc^2}
  = \frac{4(1-\nu)}{5-6\nu},
\label{gamma1smalla}
\end{equation}
where we have used eqs.\ (\ref{lambdadef}) and (\ref{lcdef}). Equation
(\ref{gamma1smalla}) describes the decrease of $\gamma$ due to network
compressibility. (See fig.\ \ref{fig_gamma_rc}(a).)  The reduction is
not sharp; for a Poisson ratio $\nu=0.4$ it amounts to less than 10\%.
In the limit of an incompressible network, for these boundary
conditions, the GSR holds for any value of $a$.

Note that the transition from eq.\ (\ref{gamma1largea}) to
eq.\ (\ref{gamma1smalla}) upon decreasing $a$ can be recast as a
transition with decreasing frequency; as $\omega$ gets smaller, the
ratio $\etab(\omega)/\eta$ increases, making eq.\ (\ref{gamma1smalla})
with its compressibility correction the valid one. This is in line
with the picture presented in ref.\ \cite{Fu2008}.

The asymptotic flow has its universal form, eq.\ (\ref{Oseen}). The
full expressions for the sub-asymptotic flow and the resulting
crossover lengths are given in ref.\ \cite{supplementary}. Assuming
that $\xi$ is the smallest length, there are two limiting cases. For a
large sphere,
\begin{eqnarray}
  a\gg\lc:\ &&i\omega\vecu^{(1)} \simeq \vecv^{(1)} \simeq
  -\frac{a^2}{24\pi\etab r^3} \left(2F_r\rhat -
  F_\theta\thetahat\right) \nonumber\\ &&\rc^{(v)}\simeq\rc^{(u)}
  \simeq a/\sqrt{3}.
\label{subasymp1largea}
\end{eqnarray} 
Thus, in this limit, also the sub-asymptotic terms describe collective
flow of the two components together. [This is only approximate,
  though, since there is always the universal relative flow of
  eq.\ (\ref{subasympVr}).]  These terms set in only in the near field
and, therefore, are not of much interest. Equations (\ref{Oseen}) and
(\ref{subasymp1largea}) are identical to the Stokes flow due to a
sphere moving in a fluid of viscosity $\etab$.  For a smaller sphere,
\begin{eqnarray}
  a\ll\lc:\ &&\vecv^{(1)} \simeq -\frac{a^2}{48(1-\nu)\pi\etab
    r^3} \left(2F_r\rhat - F_\theta\thetahat\right) \nonumber\\
  &&i\omega\vecu^{(1)} \simeq -\frac{\lc^2}{4\pi\etab
    r^3} \left(2F_r\rhat - F_\theta\thetahat\right) \nonumber\\
  &&\rc^{(v)} \simeq a/\sqrt{6(1-\nu)} \nonumber\\
  &&\rc^{(u)} \simeq \sqrt{2}\;\lc.
\label{subasymp1smalla}
\end{eqnarray}
In this limit the two components flow with different velocities. The
network flow is larger and crosses over to the asymptotic one at a
larger distance. The sign of the sub-asymptotic terms in
eqs.\ (\ref{subasymp1largea}) and (\ref{subasymp1smalla}) indicates
the dominance of the force-quadrupole contribution to these terms
\cite{prl14}, leading to a negative correction to the longitudinal
response and a positive correction to the transverse one.

\subsection{Sticking fluid and slipping network}
\label{sec_bc2}

If the network is allowed to fully slip over the sphere surface, the
four boundary conditions become
\begin{equation}
  \vecv(a,\theta) = U\zhat,\ \ 
  i\omega u_r(a,\theta) = U\cos\theta,\ \
  \sigma^{(u)}_{r\theta}(a,\theta) = 0.
\end{equation}

The friction coefficient in this case turns out as
\begin{eqnarray}
\label{gamma2}
  &&\frac{\gamma}{6\pi\etab a} = \\
  &&1 - \frac{(\lc^2-\xi^2)(a^2+2\lambda^2+2a\lambda+2\lc^2)}
  {(a^2+2\lambda^2+2\lambda a)(a\xi+3\lc^2) + 2\lc^2(a\xi+\lc^2)}.
\nonumber
\end{eqnarray}
Analysis of eq.\ (\ref{gamma2}) leads to the following three
limits. For a sufficiently large sphere, satisfying either
$a\gg\max(\lc^2/\xi,\lambda)$ or $\lc^2/\xi\ll a\ll\lambda$, we obtain
the GSR of eq.\ (\ref{gamma1largea}). This condition is even harder to
fulfill than the one in sect.\ \ref{sec_bc1}.  For a smaller sphere,
we obtain a modified relation,
\begin{equation}
  a\ll\min(\lc^2/\xi,\lambda):\ 
  \frac{\gamma}{6\pi\etab a} \simeq 
  \frac{2\lambda^2}{3\lambda^2+\lc^2} =
  \frac{4(1-\nu)}{7-8\nu}.
\label{gamma2smalla}
\end{equation}
In the limit of an incompressible network this expression describes a
sphere with slip boundary conditions in a fluid of viscosity
$\etab(\omega)$, $\gamma=4\pi\etab a$. The third limit, $\lambda\ll
a\ll\lc^2/\xi$, gives the same compressibility-independent full-slip
result. Note again that the same change from the GSR
[eq.\ (\ref{gamma1largea})] to a slip- and compressibility-dependent
relation [eq.\ (\ref{gamma2smalla})] is achieved also by decreasing
$\omega$, as noted in ref.\ \cite{Fu2008}.

The asymptotic flow remains that of eq.\ (\ref{Oseen}). We give the
full expressions for the sub-asymptotic flow and crossover lengths in
ref.\ \cite{supplementary}. Assuming again that $\xi$ is the smallest
length, there are two limiting cases. For a very large sphere,
\begin{eqnarray}
  a\gg\lambda:\ &&\vecv^{(1)} \simeq -\frac{1}{2}i\omega\vecu^{(1)}
  \simeq \frac{\lc^2}{12\pi\etab r^3}
  \left(2F_r\rhat - F_\theta\thetahat\right)
  \nonumber\\ &&\rc^{(v)}\simeq\rc^{(u)}/\sqrt{2} \simeq \sqrt{2/3}\;\lc.
\label{subasymp2largea}
\end{eqnarray} 
For a smaller sphere,
\begin{eqnarray}
  a\ll\lambda:\ &&\vecv^{(1)} \simeq \frac{a^2(1-2\nu)}{48(1-\nu)\pi\etab
    r^3} \left(2F_r\rhat - F_\theta\thetahat\right) \nonumber\\
  &&i\omega\vecu^{(1)} \simeq -\frac{\lc^2}{4\pi\etab
    r^3} \left(2F_r\rhat - F_\theta\thetahat\right) \nonumber\\
  &&\rc^{(v)} \simeq \sqrt{(1-2\nu)/[6(1-\nu)]}\;a \nonumber\\
  &&\rc^{(u)} \simeq \sqrt{2}\;\lc.
\label{subasymp2smalla}
\end{eqnarray}
Therefore, for the case of a slipping network we find in the
sub-asymptotic region relative flow of the two components. Note the
change of sign of the sub-asymptotic $\vecv^{(1)}$ in
eqs.\ (\ref{subasymp2largea}) and (\ref{subasymp2smalla}) as compared
to sect.\ \ref{sec_bc1}. This marks the dominance of the mass-dipole
effect in these $1/r^3$ terms, leading to a positive correction to the
longitudinal response and a negative correction to the transverse
one. Yet, as in sect.\ \ref{sec_bc1}, these effects are of limited
relevance as they set in at distances comparable to the particle size.

\subsection{Sticking fluid and free network}
\label{sec_bc3}

The third and last case that we consider is a free network, \ie one
that does not directly exchange stresses with the sphere. The network
moves only because of its coupling to the fluid. Discussing the
physical relevance of these conditions is deferred to
sect.\ \ref{sec_discuss}. The corresponding four boundary conditions
are
\begin{equation}
  \vecv(a,\theta) = U\zhat,\ \ 
  \sigma^{(u)}_{rr}(a,\theta)=\sigma^{(u)}_{r\theta}(a,\theta) = 0.
\end{equation}

Here, the force exerted on the sphere comes solely from the fluid
component, without contribution from the network. The friction
coefficient turns out as
\begin{eqnarray}
  \frac{\gamma}{6\pi\etab a} &=& \frac{a^2\lambda(a+\lambda)
  (a\xi +\xi^2 + 2\lc^2) +}
  {a^2\lambda(a+\lambda)(a\xi + 3\lc^2) +}\cdots
\label{gamma3}
 \\
  &\cdots& \frac{+\; 2\lc^2\xi(a+3\lambda)(2a+3\lambda)
  (a + \xi)}{+\; 2\lc^2(a+3\lambda)(2a+3\lambda)
  (a\xi + \lc^2)}.
 \nonumber
\end{eqnarray}
The condition of a free network should be appropriate, in particular,
for beads smaller than the mesh size. Hence, unlike the preceding two
sub-sections, we do not assume that $\xi$ is the smallest scale.  This
leads to three limiting cases. For a very large sphere, satisfying
either $a\gg\max(\lc^2/\xi,\lambda)$ or $\lc^2/\xi\ll a\ll \lambda$,
the GSR of eq.\ (\ref{gamma1largea}) is recovered. For smaller
spheres, where $\lc\ll a\ll\min(\lc^2/\xi,\lambda)$ or $\lambda\ll
a\ll\lc^2/\xi$, the effective-slip result is obtained, $\gamma\simeq
4\pi\etab a$. For yet smaller spheres, we get
\begin{equation}
  a\ll\lc:\ \  
  \gamma \simeq 6\pi\eta a \left[a^2/(9\xi^2)+a/\xi+1\right].
\label{gamma3smalla}
\end{equation}
For the current boundary conditions, and spheres smaller than the mesh
size, eq.\ (\ref{gamma3smalla}) enables us to recover the ordinary
Stokes relation, $\gamma\simeq 6\pi\eta a$. For spheres of
intermediate size, $\xi\ll a\ll\lc$, a different relation is obtained
from eq.\ (\ref{gamma3smalla}), $\gamma\simeq(2/3)\pi\eta
a^3/\xi^2$. Note that both of these limiting expressions are
independent of compressibility.

The asymptotic flow remains in its universal form,
eq.\ (\ref{Oseen}). The full expressions for the sub-asymptotic flows
and crossover distances are found in ref.\ \cite{supplementary}. We
present results in the following two opposite limits. For very large
spheres the sub-asymptotic flows and crossover distances coincide with
those of a sphere in an ordinary Stokes flow as found in
sect.\ \ref{sec_bc1}, eq.\ (\ref{subasymp1largea}). For small spheres,
however, we obtain
\begin{eqnarray}
  a\ll\lc:\  
  \vecv^{(1)} &\simeq& \frac{3\lc^2[a^2+3\xi(a+\xi)]}
  {4\pi\etab[a^2+9\xi(a+\xi)] r^3}
  \left(2F_r\rhat - F_\theta\thetahat\right) \nonumber\\
  i\omega\vecu^{(1)} &\simeq& \frac{\lc^2a^2}
  {2\pi\etab[a^2+9\xi(a+\xi)] r^3}
  \left(2F_r\rhat - F_\theta\thetahat\right)\nonumber\\
  \rc^{(v)} &\simeq& \left(\frac{6[a^2+3\xi(a+\xi)]}
  {a^2+9\xi(a+\xi)} \right)^{1/2} \lc \nonumber\\
  \rc^{(u)} &\simeq& \frac{2a}
  {[a^2+9\xi(a+\xi)]^{1/2}}\;\lc.
\label{subasymp3smalla}
\end{eqnarray} 
As in sect.\ \ref{sec_bc2}, the sign of the sub-asymptotic flows for
$a\ll\lc$ indicates the dominance of the mass-dipole contribution to
the $1/r^3$ terms. However, in the present case, the effect is
amplified by a much larger mass dipole, proportional to
$\lc^2/\etab=\xi^2/\eta$. This is because the bead displaces only the
local fluid as it oscillates without contact with the
network. Consequently, we expect large positive and negative
corrections, respectively, to the longitudinal and transverse
responses, pushing the crossover between the sub-asymptotic and
asymptotic behaviors further away to a distance $\sim\lc$, much larger
than both $\xi$ and $a$.

\subsection{Scaling relations}
\label{sec_scaling}

The problem which we have been studying depends on three
lengths\,---\,$a$, $\xi$, $\lambda$\,---\,and two
viscosities\,---\,$\eta$, $\etab$. (Equivalently, one of the
viscosities can be traded for the fourth length $\lc$.) Any quantity
of interest, such as $\gamma$, $r^3v^{(1)}$, or $\rc^{(u,v)}$, once
properly scaled, can be expressed as a dimensionless function of
$\xi/a$, $\lambda/a$, and $\etab/\eta$. If we assume the limits of
large $\lambda/a$ and $\etab/\eta$ (both satisfied in most practical
circumstances), we are left with functions of $\xi/a\equiv x$
alone. Such single-variable scaling functions are useful for
comparison between experiment and theory \cite{prl14} and for reliable
extraction of material parameters, such as the correlation length
$\xi$ \cite{SonnSegev14}. We concentrate on the properties $\gamma$,
$\vecv^{(1)}$, and $\rc^{(v)}$, in the limit $a\ll\lc$, as these are the
most relevant experimentally.

We begin with the sphere's friction coefficient $\gamma$, which is
directly measured in one-point microrheology. We define the scaling
function $\tilde\gamma_i \equiv \gamma_i/(6\pi\etab a)$, where the
index $i=1,2,3$ corresponds to the three sets of boundary conditions
treated in sects.\ \ref{sec_bc1}, \ref{sec_bc2}, and
\ref{sec_bc3}. From eqs.\ (\ref{gamma1smalla}), (\ref{gamma2smalla}),
and (\ref{gamma3smalla}) we find, respectively,
\begin{eqnarray}
  \tilde\gamma_1(x) &=& 4(1-\nu)/(5-6\nu) = {\rm const} \nonumber\\
  \tilde\gamma_2(x) &=& 4(1-\nu)/(7-8\nu) = {\rm const} \nonumber\\
  (\etab/\eta)\tilde\gamma_3(x) &=& [1/9+x(1+x)]/x^2.
\label{gammascaling}
\end{eqnarray}

We turn next to the sub-asymptotic flow $\vecv^{(1)}$ and focus on its
radial component at zero angle, $v^{(1)}_r(r,0)$. This function can be
measured from the longitudinal displacement correlations of particle
pairs in the extended two-point technique \cite{prl14,SonnSegev14}. We
define the scaling function $\tilde{v}^{(1)}_i \equiv
[4\pi\etab/(a^2F_r)]r^3v^{(1)}_{r,i}$. Equations
(\ref{subasymp1smalla}), (\ref{subasymp2smalla}), and
(\ref{subasymp3smalla}) yield, respectively,
\begin{eqnarray}
  \tilde{v}^{(1)}_1(x) &=& -1/[6(1-\nu)] = {\rm const} 
 \nonumber\\
  \tilde{v}^{(1)}_2(x) &=& (1-2\nu)/[6(1-\nu)] = {\rm const}
\label{tildev}\\
  (\eta/\etab)\tilde{v}^{(1)}_3(x) &=& 6x^2 [1+3x(1+x)]/[1+9x(1+x)].
   \nonumber
\end{eqnarray}

Finally, we address the crossover distance $\rc^{(v)}$. Defining
$\tilde{r}^{(v)}_{{\rm c},i} \equiv r^{(v)}_{{\rm c},i}/a$, we find from
eqs.\ (\ref{subasymp1smalla}), (\ref{subasymp2smalla}) and
(\ref{subasymp3smalla}) the simple relation
\begin{equation}
  \tilde{r}^{(v)}_{{\rm c},i}(x) = \left| \tilde{v}^{(1)}_i(x) \right|^{1/2},
\end{equation}
where the functions $\tilde{v}^{(1)}_i(x)$ are given in
eq.\ (\ref{tildev}).

In the case of the third set of boundary conditions (sticking fluid
and free network) the rescaled sub-asymptotic flow and crossover
distance, $\tilde{v}^{(1)}_3$ and $\tilde{r}^{(v)}_{{\rm c},3}$, depend on
the fluid viscosity $\eta$. As will be discussed in
sect.\ \ref{sec_discuss}, this makes comparison between the two-fluid
model and experiment problematic. We would like to replace the
solvent's $\eta$ with the effective local viscosity sensed by the
sphere, $\etal$, \eg as inferred from one-point microrheology using
the definition $\etal\equiv\gamma/(6\pi a)$. Using eq.\ (\ref{gamma3})
we make this replacement and obtain
\begin{equation}
  (\tilde{r}^{(v)}_{{\rm c},3})^2 = \tilde{v}^{(1)}_3 =
  2 (\etab/\etal) (x^2+x+1/3).
\label{expscaling}
\end{equation}

\section{Discussion}
\label{sec_discuss}

\subsection{Network--particle boundary conditions}

We have formulated the particle--medium coupling in terms of boundary
conditions at the sphere surface, without abandoning the continuum
description. While the solvent has been assumed to satisfy the stick
boundary condition, three different sets of boundary conditions have
been assumed for the network. On top of the previously studied network
stick and slip \cite{Fu2008}, we have introduced boundary conditions
corresponding to a free network that does not exchange stresses
directly with the sphere. We expect the different boundary conditions
to hold in different experimental scenarios. In cases where the
network is bound to the sphere, stick boundary conditions should
clearly hold. In scenarios where the sphere is surrounded by solvent
without direct contact with the network, the free-network boundary
conditions should be valid. This will happen, for example, for
particles much smaller than the network mesh size and for those moving
inside a solvent ``cage''. It is unclear what boundary conditions
should be used in-between these two extremes of strong contact and no
contact. For flexible polymer networks, which are neither bound to nor
depleted from the bead, the appropriate boundary conditions might be
those of full or partial slip \cite{Starrs2003}.

We have focused on the effect of these different particle--medium
couplings on the following dynamic properties: (a) the sphere's
friction coefficient $\gamma$ (equivalently, its displacement
autocorrelation in equilibrium); (b) the asymptotically far flow field
$\vecv^{(0)}$ created by the sphere's motion (equivalently, the
displacement pair correlation function at large separations); (c) the
sub-asymptotic flow field $\vecv^{(1)}$ (displacement pair correlation
function at intermediate separations). All three properties are
directly accessible by microrheology\,---\,the first through one-point
measurements, and the other two through two-point ones.  We now discuss
the results for these three properties.

\subsection{Local dynamics}

The results for $\gamma$ underline the sensitivity of this one-point
property to the immediate environment of the particle, as was
indicated by earlier studies
\cite{Crocker2000,Chen2003,Starrs2003,Fu2008,Levine2000,Levine2001b,Levine2001a}.
We find that the GSR\,---\,the assumption underlying one-point
microrheology\,---\,is generally violated. It is valid only in two
quite strict limits: (a) a very large sphere, whose size exceeds $\lc$
or $\lambda$, \ie a length proportional to the mesh size times a large
factor dependent on $\etab(\omega)/\eta$; or (b) a smaller sphere
comparable to the mesh size {\em together} with a very high frequency,
such that $\etab(\omega)\simeq\eta$. Otherwise, there are significant
deviations from the GSR, as given by eqs.\ (\ref{gamma1smalla}),
(\ref{gamma2smalla}), and (\ref{gamma3smalla}). Naturally, the
deviation is particularly large for the free-network boundary
conditions [eq.\ (\ref{gamma3smalla})], where the bulk viscosity
$\etab(\omega)$ is replaced by a local viscosity which is usually much
smaller. (Theoretically, according to eq.\ (\ref{gamma3smalla}), it
should be the solvent viscosity $\eta$; yet, this is not so in
practice. See the discussion below.)

\subsection{Far flow}

The results for $\vecv^{(0)}$, by contrast, are universal. The
asymptotic flow always obeys, exactly, a generalized Oseen tensor for
a fluid with effective shear viscosity $\etab(\omega)$,
eq.\ (\ref{Oseen}), which is independent of boundary conditions and
network compressibility. This finding is in apparent contradiction
with earlier theoretical results, which we now discuss in detail.

Levine and Lubensky observed that the component of the two-point
correlations perpendicular to the line connecting the two points, at
large separations, depended on compressibility
\cite{Levine2000,Levine2001b}. They suggested, therefore, to extract
the compressibility from the ratio of perpendicular to parallel
components of two-point measurements. The reason for the appearance of
compressibility in their asymptotic expressions was that they
considered a compressible viscoelastic bulk, whereas in the two-fluid
model there is a background of incompressible fluid. As shown above,
at large distances the network and its solvent move together; the
relevant compressibility, therefore, is that of the medium as a whole
(essentially the solvent's), which is, to a very good approximation,
zero. The suggestion made in refs.\ \cite{Levine2000,Levine2001b} was
subsequently tried experimentally \cite{Gardel2003,Pelletier2009}. A
Poisson ratio of $0.5$ (negligible compressibility) was measured for
actin networks \cite{Gardel2003,Pelletier2009}, and a slightly smaller
value for a mixture of actin and microtubules
\cite{Pelletier2009}. Thus, a reliable microrheology measurement of
the compressibility of biopolymer networks has not been achieved
\cite{MacKintosh2008}. According to our analysis [eq.\ (\ref{Oseen})]
this is simply because the $1/r$ pair correlations at large distances
do not depend at all on network compressibility; the ratio of
perpendicular to parallel components of this correlation is invariably
$1/2$.

According to Fu {\it et al.} \cite{Fu2008} the far flow depends on the
particular choice of boundary conditions at the sphere surface. This
disagreement with our results is readily resolved once we notice that
the universality of the far flow appears when we prescribe the force
$\vecF$ acting on the sphere, whereas they prescribed the sphere's
velocity. In other words, the dependence of the far flow on boundary
conditions in ref.\ \cite{Fu2008} arises only from the dependence of
$\gamma$ (\ie the single-particle force--velocity relation) on those
boundary conditions. We note that the flow velocity as a function of
force is the one related to the particles' pair mobility and, hence,
the one relevant to two-point measurements.

On the one hand, the universality of $\vecv^{(0)}$ reinforces the
robustness of asymptotic two-point microrheology for measuring bulk
shear moduli. On the other hand, the pair correlations at
asymptotically large distances cannot reveal separate properties of
the network and solvent such as network compressibility. A way to
measure the compressibility is outlined below in sect.\ \ref{sec_exp}.

\subsection{Intermediate flow}

The form of the relative flow at intermediate distances is universal
as well, assuming that one knows $\lc=(\etab/\eta)^{1/2}\xi$;
cf.\ eq.\ (\ref{subasympVr}). However, the sub-asymptotic flow
accessible to two-point experiments, $\vecv^{(1)}$, does depend on
boundary conditions. In the case of sticking or slipping network it is
of limited interest since it sets in at distances comparable to the
particle size. In the free-network case the mass-dipole effect causing
the sub-asymptotic flow is strongly enhanced, as the bulk viscosity is
replaced by the local one. This makes the $1/r^3$ corrections to the
asymptotic $1/r$ terms significant far from the sphere, making them
manifest in two-point microrheology \cite{prl14}.

\subsection{Biopolymer networks}

\begin{figure}[tbh]
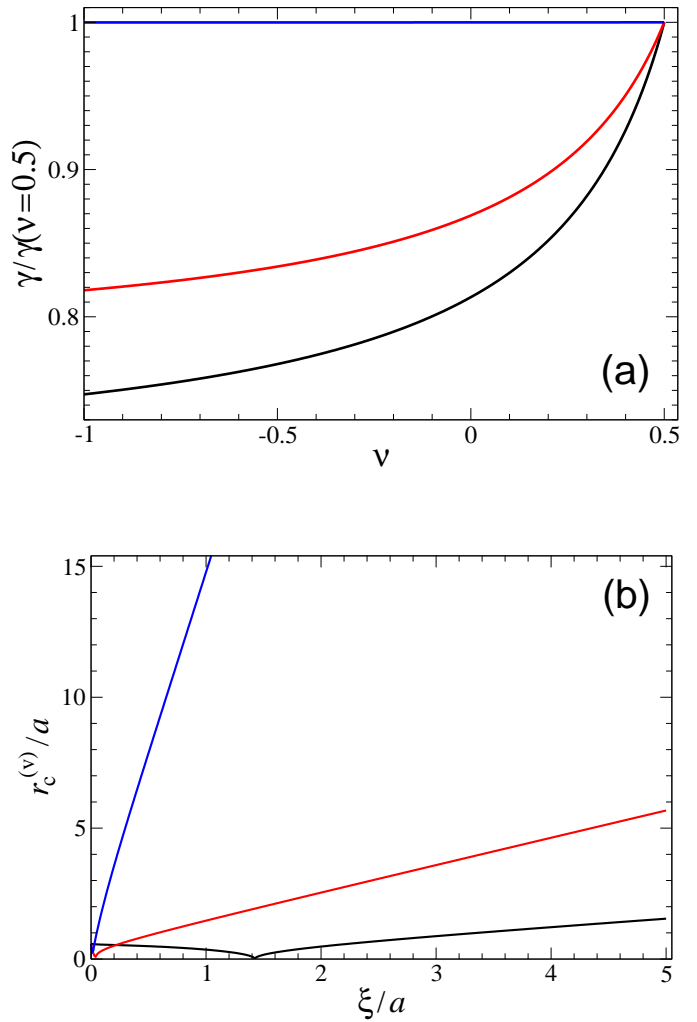

\vspace{0.6cm}
\centerline{\resizebox{0.49\textwidth}{!}
{\includegraphics{fig2a}}}
\vspace{1.1cm}
\centerline{\resizebox{0.48\textwidth}{!}
{\includegraphics{fig2b}}}
\vspace{1cm}
\caption[]{(a) Friction coefficient, scaled by its value for an
  incompressible network, as a function of network Poisson ratio. (b)
  Observable crossover distance as a function of correlation length,
  both scaled by the bead's radius. In both panels the three curves,
  from bottom to top, correspond to the three boundary conditions of
  sticking, slipping, and free network, respectively. The free-network
  case is practically independent of network compressibility (a) and
  characterized by a large crossover distance (b). Parameters:
  $\etab/\eta=100$, $\xi/a=1$ (a), $\nu=0.45$ (b).}
\label{fig_gamma_rc}
\end{figure}

The results for the three sets of boundary conditions, therefore,
differ considerably. In particular, the free-network case strongly
deviates from those of sticking and slipping networks. This is
demonstrated in fig.\ \ref{fig_gamma_rc} for the dependence of the
friction coefficient $\gamma$ on network compressibility, and the
dependence of the observable crossover distance $\rc^{(v)}$ on the
mesh size $\xi$.

In the special case of entangled F-actin networks, the appropriate
boundary conditions, even for spheres significantly larger than the
mesh size, are those of a free network. The strong evidence for this
unintuitive result comes from agreement with experiments
\cite{prl14,SonnSegev14} concerning the large prefactor of
$\etab(\omega)/\eta$, which enhances the intermediate flow
$\vecv^{(1)}$ and squared crossover length $(\rc^{(v)})^2$ compared to
the other boundary conditions. [See eqs.\ (\ref{subasymp3smalla}) and
  (\ref{tildev}).]  Given the large differences in the results between
the free-network conditions and the other two sets (see
fig.\ \ref{fig_gamma_rc}, for example), there is no way that stick or
slip boundary conditions could fit the experimental data.  The same
seems to hold for microtubule networks, where a crossover length much
larger than the mesh size was observed as well
\cite{Pelletier2009}. As mentioned above, the free-network boundary
condition leads also to the largest downward deviation of $\gamma$
from the GSR. This accounts for the orders-of-magnitude discrepancies
found between one- and two-point (or macroscopic) measurements in
actin \cite{Crocker2000,Chen2003,Gardel2003,Starrs2003} and
microtubule \cite{Pelletier2009} networks, as well as the sensitivity
of one-point measurements to the surface chemistry of the beads
\cite{McGrath2000,Valentine2004}. It may be related also to the
unexpectedly deep penetration of inert beads into actin networks
\cite{McGrath2000}.

In addition to the insensitivity of $\vecv^{(0)}$ to network
compressibility, under the free-network boundary conditions the
one-point $\gamma$ is also practically independent of compressibility
[fig.\ \ref{fig_gamma_rc}(a)]. We are led to the conclusion that
measuring this parameter for biopolymer networks using inert beads is
probably impossible.

One should be able to reproduce the results obtained above for the
case of free-network boundary conditions by introducing local
heterogeneity. For example, extending the work of
refs.\ \cite{Levine2000,Levine2001b}, one can consider a finite shell
of solvent surrounding the bead and apply the two-fluid model outside
the shell. This, however, adds another length parameter to the model,
which does not seem necessary to account for experiments.

The physical origin of the ``phantom network'' condition\,---\,the
lack of contact between a semiflexible network and a bead larger than
its mesh size\,---\,remains unclear. A depletion layer resulting from
sphere--network repulsion, be it a result of a genuine interaction
potential or an entropy-induced effect, would not prevent direct
exchange of stresses over the experimentally relevant time scales. For
dilute semiflexible networks that do not bind to the beads these
stresses are apparently inconsequential.

\subsection{The intermediate dynamic length}

The entire analysis highlights the key role played by an intermediate
dynamic length, $\lc(\omega)\equiv[\etab(\omega)/\eta]^{1/2}\xi$,
which was first pointed out in ref.\ \cite{prl14}. It is the crossover
distance separating the intermediate, relative (Brinkman) flow from
the asymptotic, collective (Stokes) flow. Unless the frequency is very
high such that $\etab(\omega)\simeq\eta$, $\lc$ is much larger than
the mesh size $\xi$. This length affects various dynamic
properties. For example, for the GSR to hold, we have found that the
bead must be larger than $\lc$, not $\xi$. This is true for all
boundary conditions studied, even in the case of a sticking network.
(The bead should be even larger, $a\gg\lc^2/\xi$ in the cases of
slipping or free networks.) This underlines again the problem in using
one-point microrheology based on the GSR or GSER. The distance $\lc$,
although always affecting the crossover between the relative and
collective flows, may be masked by a mixture of terms when the
crossover distance of the observable flow, $\rc^{(v)}$, is
measured. However, in the case of free-network boundary conditions, as
in actin networks, $\rc^{(v)}$ is proportional to $\lc$, pushing the
observed crossover to a distance much larger than $\xi$, as
demonstrated in fig.\ \ref{fig_gamma_rc}(b), and as confirmed
experimentally \cite{prl14,SonnSegev14}.

\subsection{Further experimental consequences}
\label{sec_exp}

The scaling relations derived in sect.\ \ref{sec_scaling} suggest ways
to sensitively extract various elusive parameters through extended
microrheology combining one- and two-point measurements. For example,
to get the network's Poisson ratio (which may be frequency-dependent),
one should use a bead with surface groups that bind the network
\cite{McGrath2000,Valentine2004}, to ensure stick boundary conditions,
for which the sensitivity of $\gamma$ to $\nu$ is maximum
[fig.\ \ref{fig_gamma_rc}(a)]. One may then apply the first relation
in eq.\ (\ref{gammascaling}) to extract $\nu$. Here, the one-point and
two-point measurements are needed to get $\gamma$ and a reliable
$\etab(\omega)$, respectively. Another example is the ability to
extract the network's correlation length and its dependence on
parameters other than network concentration, using
eq.\ (\ref{expscaling}). This has already been successfully
demonstrated in refs.\ \cite{prl14,SonnSegev14}.

\subsection{Open issues}

We conclude with three broader issues raised by the present study.

The first problem concerns a deficiency in the two-fluid model. There
is a certain inconsistency in the way the model is constructed. On the
one hand, it should break down at length scales smaller than the mesh
size $\xi$, where, obviously, the network cannot be treated as a
viscoelastic continuum. On the other hand, one of the model parameters
used at all scales is the solvent viscosity $\eta$, although it is
physically meaningful only at scales smaller than $\xi$. In the case
of the free-network boundary conditions (sect.\ \ref{sec_bc3}), the
model seems to correctly cover the entire range from
$a/\xi\rightarrow\infty$ down to $a/\xi\rightarrow 0$;
eq.\ (\ref{gamma3}) appropriately reproduces $\gamma=6\pi\etab a$ in
the former limit, and $\gamma=6\pi\eta a$ in the latter.  The validity
question can be posed in the following practical terms: can one
extract the solvent viscosity by tracking the motion of a bead larger
than $\xi$?  If one believes the expressions derived in
sect.\ \ref{sec_particular}, the answer is evidently positive; for
example, to get $\eta$ one could fit one-point measurements in actin
networks to eq.\ (\ref{gamma3}), or two-point measurements to
eq.\ (\ref{subasymp3smalla}). In practice, attempts to do so yielded
unreasonable results \cite{private}. At least in this restricted sense
of the meaning of $\eta$, the failure of the model extends to lengths
(\eg particle sizes) much larger than $\xi$ and, therefore, is
unrelated to the breakdown of the continuum limit. When using a
combination of one- and two-point measurements, there is a way to
bypass this problem, which was demonstrated to work well in the case
of actin networks \cite{prl14,SonnSegev14}. As already mentioned in
sect.\ \ref{sec_scaling}, one can use the one-point measurement to
define a local viscosity as sensed by the bead,
$\etal\equiv\gamma/(6\pi a)$, and substitute it for $\eta$ in the
expressions relevant to two-point measurements, such as $\vecv^{(1)}$
and $\rc^{(u,v)}$. Since the one-point property $\gamma$ depends on
particle size and frequency, so does the local viscosity,
$\etal=\etal(a/\xi,\omega)$, with the two limits
$\etal(0,\omega)=\eta$ and $\eta(\infty,\omega)=\etab(\omega)$.  The
necessity to define the length-scale-dependent $\etal$ indicates that
an actual complex fluid is not fully characterized by uniform moduli
as assumed by the two-fluid model. Instead, one should introduce
wavevector-dependent viscosities.

The second issue has to do with the emergence of the mass-dipole term
governing the relative, intermediate flow. The mass dipole in
eq.\ (\ref{subasympVr}), proportional to $\lc^2/\etab=\xi^2/\eta$ is
independent of the particle size.  This is curious, because a
vanishingly small particle obviously cannot displace any fluid
mass. The way in which an effective mass dipole builds up within a
correlation ``pore'' of characteristic size $\xi$, and what happens
when there are many such characteristic lengths, or none at all (as in
a fractal structure), are interesting questions to be answered.

Finally, we would like to point out a possible relation between the
dynamic length $\lc$ discussed here and the divergent dynamic length
in polymer gelation and colloidal glass transitions (and, perhaps,
amorphous solidification in general \cite{glass}). We notice that, as
the material solidifies and $\etab(\omega\rightarrow 0)$ diverges, so
does $\lc$. Within the description laid out here, the physics
accompanying this transition is intuitively clear. As the crossover
length $\lc(\omega\rightarrow 0)$ diverges, the intermediate, Brinkman
region stretches out to infinity. The fluid, flowing relative to the
network at all distances, loses its translational invariance, and the
network turns into a solid porous matrix.


\begin{acknowledgement}
I am indebted to Adar Sonn-Segev and Yael Roichman for a fruitful
collaboration and many discussions. Helpful discussions with Yitzhak
Rabin and Tom Witten are gratefully acknowledged. This research was
supported by the Israel Science Foundation under Grants No.\ 8/10 and
No.\ 164/14.
\end{acknowledgement}

\end{document}